\documentclass[twocolumn,pra,showpacs,superscriptaddress,amssymb,amsmath]{revtex4-1}
\usepackage{graphicx}
\usepackage{epstopdf}
\usepackage{bm}
\usepackage[colorlinks=true,linkcolor=blue,urlcolor=blue,citecolor=blue]{hyperref}
\usepackage{comment}
\usepackage{float}
\usepackage{cancel}
\usepackage{times}
\usepackage{xcolor}

\begin{document}

\title{Chemical reactions of ultracold alkaline-earth-metal diatomic molecules}

\author{Hela Ladjimi}
\affiliation{Faculty of Physics, University of Warsaw, Pasteura 5, 02-093 Warsaw, Poland}
\author{Micha{\l} Tomza}
\email{michal.tomza@fuw.edu.pl}
\affiliation{Faculty of Physics, University of Warsaw, Pasteura 5, 02-093 Warsaw, Poland}

\date{\today}

\begin{abstract}
We study the energetics of chemical reactions between ultracold ground-state alkaline-earth-metal diatomic molecules. We show that the atom-exchange reactions forming homonuclear dimers are energetically allowed for all heteronuclear alkaline-earth-metal combinations. We perform high-level electronic structure calculations on the potential energy surfaces of all possible homo- and heteronuclear alkaline-earth-metal trimers and show that trimer formation is also energetically possible in collisions of all considered dimers. Interactions between alkaline-earth-metal diatomic molecules lead to the formation of deeply bound reaction complexes stabilized by large non-additive interactions. We check that there are no barriers to the studied chemical reactions. This means that all alkaline-earth-metal diatomic molecules are chemically unstable at ultralow temperature, and optical lattice or shielding schemes may be necessary to segregate the molecules and suppress losses.
\end{abstract}

\maketitle

\section*{Introduction}
The achievement of quantum degeneracy in ultracold gases of alkaline-earth-metal atoms~\cite{FukuharaPRA2007,StellmerPRL09,KraftPRL09,DeSalvoPRL10} and their application in precision measurements~\cite{LudlowRMP15,SafronovaRMP18} and quantum-many body physics~\cite{CazalillaRPP14,ZhangNRP20} have paved the way for the production of ultracold alkaline-earth-metal diatomic molecules. Such molecules can be formed using photoassociation followed by an optical stabilization~\cite{ReinaudiPRL12,LeungNJP21} or the stimulated Raman adiabatic passage  from atom pairs on sites of an optical lattice~\cite{StellmerPRL12}. The association using optical Feshbach resonances~\cite{EnomotoPRL08} may also be possible, however, less efficient than the magnetoassociation of alkali-metal atoms via magnetic Feshbach resonances~\cite{JulienneRMP06b}. Ultracold alkaline-earth-metal diatomic molecules offer numerous exciting research prospects in quantum physics and chemistry, ranging from studying quantum-controlled chemical reactions~\cite{McdonaldNature16} and quantum simulations~\cite{BhongalePRL13} to precision measurements probing the fundamental laws of nature~\cite{ZelevinskyPRL08,KotochigovaPRA2009,McGuyerNP15} and metrology with molecular clocks~\cite{BorkowskiPRL18,KondovNP19}.  

Recently, fast chemical reactions with rate constants close to the universal limit~\cite{IdziaszekPRL10} were experimentally observed between ultracold ground-state Sr$_2$ molecules~\cite{LeungNJP21}. These unexpected reactive losses may jeopardize some of the above-mentioned applications in precision physics. Therefore, understating the chemical reactivity of ultracold alkaline-earth-metal diatomic molecules in the ground $X^1\Sigma^+$ electronic state is an important research goal that, to the best of our knowledge, has not been studied theoretically yet. In general, the following bimolecular reaction processes may lead to collisional losses: 
\begin{eqnarray}
AB+AB &\rightarrow& A_2 + B_2\,, \label{eq:R1} \\
AB + AB &\rightarrow& A_2B + B\,, \label{eq:R2}
\end{eqnarray}
where the reaction~\eqref{eq:R1} is the atom-exchange chemical reaction (bond-swapping reaction between polar heteronuclear reactants leading to nonpolar homonuclear products) and the reaction~\eqref{eq:R2} is the atom-transfer chemical reaction (trimer-formation reaction). Photo- or collision-induced losses via the formation of long-lived four-atom complexes~\cite{ChristianenPRL19,JachymskiPRA22,Man2022,BauseJPCA23} may also have a similar density scaling, but they are out of the scope of this work.

The reactions~\eqref{eq:R1} and~\eqref{eq:R2} between ground-state alkali-metal diatomic molecules were studied theoretically~\cite{ZuchowskiPRA10,ByrdPRA10,TomzaPRA13} and experimentally~\cite{OspelkausScience10,HuScience19}. The atom-exchange reactions are energetically allowed for some $X^1\Sigma^+$ singlet-state~\cite{ZuchowskiPRA10} and almost-all $a^3\Sigma^+$ triplet-state~\cite{TomzaPRA13} alkali-metal dimers, while the trimer formation reactions are energetically forbidden for all singlet-state~\cite{ZuchowskiPRA10} and allowed for all triplet-state~\cite{TomzaPRA13} ones. A similar pattern of chemical reactivity was also predicted for isoelectronic molecules containing silver and copper atoms~\cite{SmialkowskiPRA21}. Chemical reactions between $X^2\Sigma^+$ doublet-state alkali-metal--alkaline-earth-metal diatomic molecules are less studied, but recently ultracold RbSr molecules were theoretically predicted to be chemically reactive~\cite{ManNJP22} with energetically allowed and barrierless both atom-exchange and atom-transfer reactions on both singlet and triplet intermolecular potential energy surfaces. A similar pattern of chemical reactivity was also predicted for isoelectronic $X^2\Sigma^+$-state YbCu, YbAg, and YbAu molecules~\cite{TomzaNJP21}.

To assess the energetics of the reactions~\eqref{eq:R1} and~\eqref{eq:R2}, a comprehensive knowledge of the dissociation energies of diatomic and triatomic molecules is needed. Detailed experimental spectroscopic studies are available for a few homonuclear alkaline-earth-metal dimers~\cite{MerrittScience09,BalfourCJP70,AtmanspacherCP85,AllardEPJD03,SteinEPJD10}, while others have been studied theoretically only (see Ref.~\cite{LadjimiPRA22} and references therein). Homonuclear alkaline-earth-metal triatomic and larger clusters have been extensively studied theoretically 
\cite{KaplanJCP2000,KaplanMP2002,KlosJCP08,DiazJMS2010,AmaroJCP11,DiazCP2011,RamirezIJQC12,KalemosJCP16, MatoJCP2022} and have been demonstrated to be stabilized by non-additive many-body interactions, similar to spin-polarized alkali-metal clusters~\cite{SoldanPRA03,SoldanPRA08,SoldanPRA10,TomzaPRA13}.

\begin{table*}[t!]
\caption{The dissociation energies $D_0$ (in cm$^{-1}$) for alkaline-earth-metal diatomic molecules $AB$ in the lowest ro-vibrational level ($v=0,j=0$) of the $X^1\Sigma^+$ ground electronic state.\label{tab:D0}}
\begin{ruledtabular}
\begin{tabular}{lrrrrrr}
$A \,\backslash \,B$ & Be & Mg & Ca & Sr & Ba & Ra \\
\hline
Be & 806.5(5)~\cite{MerrittScience09} &  394~\cite{LadjimiPRA22}&726~\cite{LadjimiPRA22}& 705~\cite{LadjimiPRA22} &  815~\cite{LadjimiPRA22}  &591~\cite{LadjimiPRA22} \\
Mg &    & 405.1(5)~\cite{BalfourCJP70}  & 661.9~\cite{AtmanspacherCP85}& 650~\cite{LadjimiPRA22}  &732~\cite{LadjimiPRA22}& 592~\cite{LadjimiPRA22} \\
Ca &    &    & 1069.87(1)~\cite{AllardEPJD03} & 1020~\cite{LadjimiPRA22} & 1170~\cite{LadjimiPRA22}&  922~\cite{LadjimiPRA22} \\
Sr &    &    &    &  1061.58(1)~\cite{SteinEPJD10}  & 1172~\cite{LadjimiPRA22}& 930~\cite{LadjimiPRA22} \\
Ba &    &    &    &    &  1349~\cite{LadjimiPRA22} & 1055~\cite{LadjimiPRA22} \\
Ra &    &    &    &    &    &  848~\cite{LadjimiPRA22} \\
\end{tabular} 
\label{table:dissociation-energy}
\end{ruledtabular}
\end{table*}

In this Letter, we investigate the energetics of chemical reactions between ultracold alkaline-earth-metal diatomic molecules in their ro-vibrational and electronic ground states. To this end, we calculate the equilibrium geometries and dissociation energies of all triatomic molecules $A_2B$ ($A,B$ = Be, Mg, Ca, Sr, Ba, Ra), which are stabilized by strong non-additive three-body interactions. We show that both atom-exchange and trimmer-formation reactions are energetically allowed and barrierless for all alkaline-earth-metal diatomic molecules. We discuss implications for ultracold molecular physics experiments. Our findings explain recent experimental observations of fast reactive trap losses between ultracold ground-state Sr$_2$ molecules~\cite{LeungNJP21}. 

\section*{Electronic structure calculations}

We collect the available experimental~\cite{MerrittScience09,BalfourCJP70,AtmanspacherCP85,AllardEPJD03,SteinEPJD10} or our recent theoretical~\cite{LadjimiPRA22} dissociation energies $D_0$ for all alkaline-earth-metal diatomic molecules in the lowest ro-vibrational level of the $X^1\Sigma^+$ ground electronic state in Table~\ref{tab:D0}~{\footnote{{See Supplemental Material at https://journals.aps.org/pra/supplemental/10.1103/PhysRevA.108.L021302 for Tables I, II, and IV recalculated using our recent theoretical data~\cite{LadjimiPRA22} only.}}}. Theoretical values are calculated from the well depths $D_e$ and harmonic constants $\omega_e$ as $D_0\approx D_e-\omega_e/2$~{\footnote{{This approximation introduces errors smaller than 1$\,$cm$^{-1}$ for all molecules except ones containing Be.}}}. The theoretical results were obtained using high-level \textit{ab initio} electronic structure theory: the closed-shell coupled-cluster method restricted to single, double, and noniterative triple excitations, CCSD(T)~\cite{MusialRMP07}, with the large aug-cc-pwCV5Z Gaussian basis sets, corrected by including the full iterative triple-excitation correction from the CCSDT method with the aug-cc-pwCVTZ basis sets and full iterative quadruple-excitation correction from the valence CCSDTQ method with the aug-cc-pVTZ basis sets. The scalar relativistic effects in heavier atoms were included by employing the small-core relativistic energy-consistent pseudopotentials. More details are described in Ref.~\cite{LadjimiPRA22}.

We investigate the potential energy surfaces (PESs) of all possible homo- and heteronuclear alkaline-earth-metal triatomic molecules (and selected tetramers) in their ground electronic states using the closed-shell coupled-cluster method with the single, double, and noniterative triple excitations, CCSD(T). We employ the augmented correlation-consistent polarized weighed core-valence quintuple-$\zeta$  quality  (aug-cc-pwCV5Z) basis sets for  Be and Mg atoms~\cite{PrascherTCA10}. The scalar relativistic effects in heavier atoms are included by employing the small-core relativistic energy-consistent pseudopotentials (ECP$n$MDF) to replace $n$ inner-shell electrons~\cite{DolgCR12}. ECP10MDF, ECP28MDF, ECP46MDF, and ECP78MDF~\cite{LimJCP06} are used for Ca, Sr, Ba, and Ra, respectively, with the associated aug-cc-pwCV5Z-PP basis sets~\cite{HillJCP17}. We correct the basis set superposition error by using the counterpoise correction of Boys and Bernardi~\cite{BoysMP70}, 
\begin{eqnarray}\label{eq:SM}
V_{AB}&=&E_{AB}-(E_A+E_B)\,,
\\
V_{ABC}&=&E_{ABC}-(E_A+E_B+E_C)\,,\\
V_{ABCD}&=&E_{ABCD}-(E_A+E_B+E_C+E_D)\,,
\end{eqnarray}
where $E_i$, $E_{AB}$, $E_{ABC}$, and $E_{ABCD}$ are total energies of monomers, dimers, trimers, and tetramers computed in the respective multimer basis sets at different geometries.

To find equilibrium geometries for triatomic molecules $A_2B$, we explore three-dimensional potential energy surfaces around their global minima. First, we confirm that the studied molecules have their ground electronic states of the $^1A_1$ symmetry and global minima at the isosceles triangular geometry within the $C_{2v}$ point group. Homonuclear trimers $A_3$ additionally show threefold rotational symmetry. Next, we assume the isosceles triangular geometry with each of the $A$ atoms bound to the $B$ atom situated on the symmetry axis of the molecule. Therefore the PESs become two-dimensional functions of two coordinates $V_{A_2B}(R_,\theta)$, where $R$ is the distance between the $B$ atom and each of the $A$ atoms, and $\theta$ is the angle between the two legs of the triangle. We calculate the PESs spanned on a grid of 25-100 points being products of 5-10 interatomic distances around corresponding dimers' equilibrium distances and 5-10 angles around 60$^\circ$~\cite{SmialkowskiPRA20}. Equilibrium geometries are found by the minimalization of numerically interpolated PESs.

The interactions energies in tri- and four-atomic molecules, $V_{ABC}$ and $V_{ABCD}$, can be decomposed into the sums of pairwise additive two-body, $V_{AB}^{2b}\equiv V_{AB}$, and pairwise non-additive three-body, $V_{ABC}^{3b}$, and four-body, $V_{ABCD}^{4b}$, contributions 
\begin{eqnarray}
V_{ABC}&=&\sum_{i<j}V_{ij}^{2b}+V_{ABC}^{3b}\,,\label{eq:ABC}\\
V_{ABCD}&=&\sum_{i<j}^{}V_{ij}^{2b}+\sum_{i<j<k}^{}V_{ijk}^{3b}+V_{ABCD}^{4b}\,.\label{eq:ABCD}
\end{eqnarray} 
Combining Eqs.~\eqref{eq:SM}-\eqref{eq:ABCD} gives expressions for the non-additive three-body and four-body interactions
\begin{eqnarray}
V^{3b}_{ABC}&=&E_{ABC}-\sum_{i<j}E_{ij}+\sum_i E_i\,,\\
V^{4b}_{ABCD}&=&E_{ABCD}-\sum_{i<j<k}E_{ijk}+\sum_{i<j}E_{ij}-\sum_{i}E_i\,,
\end{eqnarray}
where all energies are computed in the respective trimers and tetramers basis sets at different geometries. 

All electronic structure calculations are performed using the \textsc{Molpro} package of $ab$ $initio$ programs~\cite{WernerJCP20}.

\begin{table}
\caption{The energy changes $\Delta E$ (in cm$^{-1}$) for the reactions $2AB\to A_2+B_2$ of alkaline-earth-metal diatomic molecules in the lowest ro-vibrational level of the $X^1\Sigma^+$ ground electronic state. \label{tab:dE}}
\begin{ruledtabular}
\begin{tabular}{lrrrrr}
$A \,\backslash \,B$  & Mg & Ca & Sr & Ba & Ra   \\
\hline
Be & -424 & -425 & -458 & -525 & -472 \\
Mg &      & -151 & -167 & -290 &  -89 \\
Ca &      &      &  -92 &  -78 &  -74 \\
Sr &      &      &      &  -66 &  -49 \\
Ba &      &      &      &      &  -87 \\
\end{tabular}
\end{ruledtabular}
\end{table}

\section*{Atom-exchange reactions}

The energy change $\Delta E$ of the atom-exchange chemical reaction~\eqref{eq:R1} is defined as a difference between the sum of dissociation energies of the heteronuclear reactants $AB$, $D_0(AB)$, and the sum of dissociation energy of the homonuclear products $A_2$ and $B_2$, $D_0(A_2)$ and $D_0(B_2)$,
\begin{equation}
\Delta E=2D_0(AB)-D_0(A_2)-D_0(B_2)\,.
\end{equation}
For $\Delta E>0$, the chemical reaction is endothermic (energetically forbidden at ultracold temperatures) and for $\Delta E<0$ -- exothermic (energetically allowed).  

Using the dissociation energies of diatomic molecules presented in Table~\ref{table:dissociation-energy}, we calculate $\Delta E$ for all alkaline-earth-metal heteronuclear combinations and collect them in Table~\ref{tab:dE}. We find that the atom-exchange chemical reactions are exothermic for all considered diatomic molecules, which can be a subject of reactive bimolecular trap losses. The released energies are between 49$\,$cm$^{-1}$ for SrRa+SrRa$\to$Sr$_2$+Ra$_2$ and 525$\,$cm$^{-1}$ for BeBa+BeBa$\to$Be$_2$+Ba$_2$ and decrease with increasing atomic masses. They are the largest for molecules containing Be due to unusually strong binding in Be$_2$. The present reaction energy changes are of the same order of magnitude as for singlet-state alkali-metal dimers~\cite{ZuchowskiPRA10} and significantly larger than for spin-polarized triplet-state alkali-metal dimers~\cite{TomzaPRA13}. We estimate the uncertainty of the calculated energy changes to be their fraction, partially due to the cancellation of correlated errors~\cite{LadjimiPRA22}. Neglected hyperfine structure and isotope shifts are also negligibly small. 

We numerically scan full potential energy surfaces and confirm for a few selected alkaline-earth-metal combinations that the formation of four-atom intermediate complexes and their dissociation proceed without any electronic barriers, similarly as for alkali-metal molecules~\cite{ZuchowskiPRA10,TomzaPRA13}. Two approaching alkaline-earth-metal molecules attract each other by a long-range dispersion interaction because they are easily polarizable~\cite{LadjimiPRA22}. Next, a relatively weak van der Waals binding with zero chemical bond order between alkaline-earth-metal atoms can easily be rearranged without forming any barriers.

The data collected in Table~\ref{table:dissociation-energy} also allows for calculating the energy changes of atom-exchange chemical reactions in atom-molecule mixtures, $AB+A \rightarrow A_2 + B$. These reactions can be either endothermic or exothermic, depending on the atomic combination.

 
\begin{table}
\caption{Characteristics of alkaline-earth-metal triatomic molecules $A_2B$ in the singlet $^1A_1$ electronic ground state. All molecules have an isosceles triangular equilibrium geometry within the $C_{2v}$ point group. Consecutive columns list: equilibrium angle between the two even legs of the triangle $\alpha_e\,$, equilibrium leg length $R_e$, well depth of the triatomic molecule $D_e$, and non-additive three-body part of the binding energy $D_e^{3b}$.\label{tab:A2B}} 
\begin{ruledtabular}
\begin{tabular}{lrrrr}
$A_2B$ & $\alpha_e\,$(degrees) & $R_e\,$(bohr) & $D_e\,$(cm$^{-1}$) & $D_e^{3b}\,$(cm$^{-1}$)\\
\hline
Be$_3$    &  60.0  & 4.15 & 8569 & 8104\\
Be$_2$Mg  &  49.1  &  5.07 & 5570& 6223 \\
Be$_2$Ca  &  43.5  &  5.62 & 7379& 7423\\
Be$_2$Sr  &  41.1  &  5.93 & 7395 & 7592\\
Be$_2$Ba  &  39.1  &  6.15 & 8635 & 8977\\
Be$_2$Ra  &  38.0  &  6.38 & 7118 & 7829\\
\hline
Mg$_2$Be  &  71.4  &  5.20 & 3305& 4024\\
Mg$_3$    &  60.0  &  6.32 & 2170& 2091\\
Mg$_2$Ca  &  54.1  &  6.79 & 3125& 2742\\
Mg$_2$Sr  &  51.4  & 7.12  & 3144& 2733\\
Mg$_2$Ba  &  49.0  & 7.29  & 3731& 3501\\
Mg$_2$Ra  &  47.8  & 7.61  & 2965& 2701\\
\hline
Ca$_2$Be  &  78.5  & 5.71  & 5947& 5378\\
Ca$_2$Mg  &  67.1  & 6.71  & 4145& 3158\\
Ca$_3$    &  60.0  & 7.37  & 5143& 3306\\
Ca$_2$Sr  &  57.1  & 7.72  & 5085& 3221\\
Ca$_2$Ba  &  54.5  & 7.96  & 5819& 3864\\
Ca$_2$Ra  &  53.2  & 8.24  & 4796& 3117\\
\hline
Sr$_2$Be  &  81.7  & 6.04  & 5788& 5299\\
Sr$_2$Mg  &  70.4  & 7.04  & 4103& 3066\\
Sr$_2$Ca  &  63.1  & 7.73  & 5010& 3100\\
Sr$_3$    &  60.0  & 8.10  & 4923& 2949\\
Sr$_2$Ba  &  57.3  & 8.35  & 5584& 3504\\
Sr$_2$Ra  &  56.0  & 8.64  & 4618& 2782\\
\hline
Ba$_2$Be  &  85.9  & 6.17  & 7850& 7540\\
Ba$_2$Mg  &  74.3  & 7.12  & 5572& 4658\\
Ba$_2$Ca  &  66.2  & 7.88  & 6494& 4393\\
Ba$_2$Sr  &  63.0  & 8.26  & 6314& 4128\\
Ba$_3$    &  60.0  & 8.54  & 7088& 4776\\
Ba$_2$Ra  &  58.7  & 8.81  & 5875& 3842\\
\hline
Ra$_2$Be  &  85.7  & 6.53 & 5059& 5118\\
Ra$_2$Mg  &  74.7  & 7.56 & 3621& 2819\\
Ra$_2$Ca  &  67.2  & 8.28 & 4409& 2774\\
Ra$_2$Sr  &  64.1  & 8.66 & 4309& 2589\\
Ra$_2$Ba  &  61.3  & 8.92 & 4846& 3039\\
Ra$_3$    &  60.0  & 9.23 & 4004& 2341\\
\end{tabular}
\end{ruledtabular}
\end{table}

\section*{Trimer-formation reactions} 

The energy change $\Delta E'$ of the trimer-formation chemical reaction~\eqref{eq:R2} is defined as a difference between the sum of dissociation energies of the heteronuclear reactants $AB$, and the dissociation energy of the triatomic product $A_2B$, $D_0(A_2B)$,
\begin{equation}
\Delta E'=2D_0(AB)-D_0(A_2B)\approx 2D_e(AB)-D_e(A_2B)\,,
\end{equation} 
where we approximate the dissociation energies by the corresponding well depths for simplicity. 

\begin{table}
\caption{The energy changes $\Delta E'$  (in cm$^{-1}$) for the reactions 2AB $\to$ A$_2$B+B of alkaline-earth-metal diatomic molecules in the lowest ro-vibrational level of the $X^1\Sigma^+$ ground electronic state.
\label{tab:dE_trimer}}
\begin{ruledtabular}
\begin{tabular}{lrrrrrr}
$A \,\backslash \,B$ & Be & Mg & Ca & Sr & Ba & Ra   \\
\hline
Be & -6700 & -4700 & -5813 & -5883 & -6899 & -5854\\
Mg & -2435 & -1310 & -1741 & -1792 & -2215 & -1737\\
Ca & -4381 & -2761 & -2939 & -2993 & -3427 & -2910\\
Sr & -4276 & -2751 & -2918 & -2759 & -3202 & -2728\\
Ba & -6114 & -4056 & -4102 & -3932 & -4356 & -3739\\
Ra & -3795 & -2393 & -2523 & -2419 & -2710 & -2288\\
\end{tabular}
\end{ruledtabular}
\end{table}

The computed equilibrium geometries and dissociation energies of triatomic alkaline-earth-metal molecules in the ground electronic state are collected in Table~\ref{tab:A2B}. We use them to calculate $\Delta E'$ for all alkaline-earth-metal homonuclear and heteronuclear combinations, which are presented in Table~\ref{tab:dE_trimer}. We find that the trimer-formation chemical reactions are highly exothermic for all considered diatomic molecules and can additionally contribute to reactive bimolecular trap losses. In the case of homonuclear dimers, these reactions are the main reactive loss channel. The released energies are between 1310$\,$cm$^{-1}$ for Mg$_2$+Mg$_2\to$Mg$_3$+Mg and 6899$\,$cm$^{-1}$ for BeBa+BeBa$\to$Be$_2$+Ba$_2$ and there is no clear trend with atomic size. We estimate the uncertainty of the calculated energy changes to be their small fraction, despite neglecting vibrational zero-point energies and less accurate description of trimers than dimers, because of very large values of the released energies. The present reaction energy changes are significantly larger than for spin-polarized triplet-state alkali-metal dimers~\cite{TomzaPRA13}, while trimer-formation reactions are energetically forbidden between ultracold singlet-state alkali-metal dimers~\cite{ZuchowskiPRA10}.

We also numerically confirm for a few selected alkaline-earth-metal combinations that the trimer-formation chemical reactions proceed without any electronic barriers, similarly to the atom-exchange ones.

\section*{Non-additive interactions}

\begin{figure*}[tb!]
\begin{center}
\includegraphics[width=0.95\textwidth]{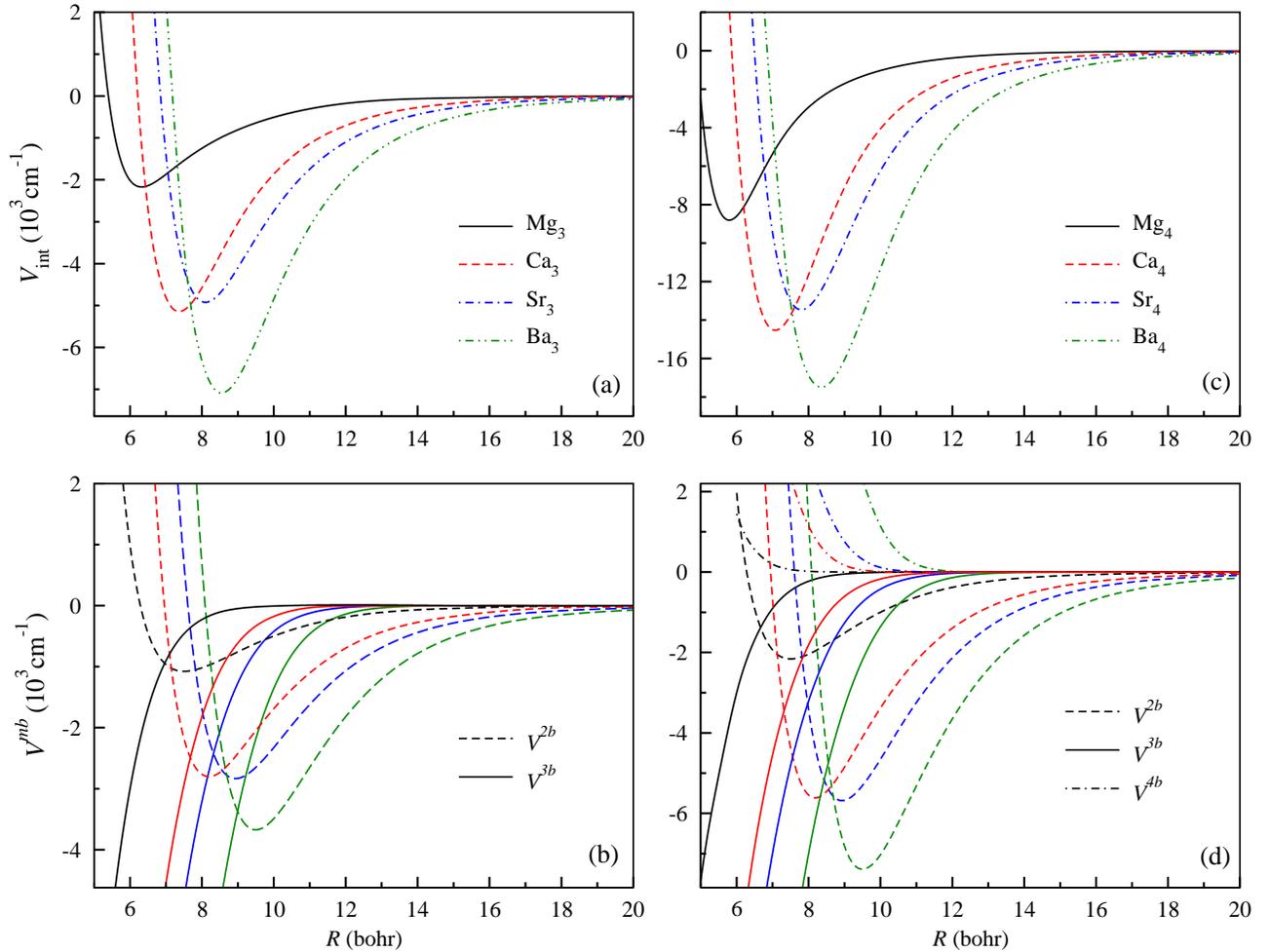}
\caption{One-dimensional cuts through the ground-state potential energy surfaces of homonuclear alkaline-earth-metal triatomic and four-atomic molecules. Panels (a) and (b) presents the results for the Mg$_3$, Ca$_3$, Sr$_3$, and Ba$_3$ trimers at the equilateral triangular geometry, while panels (c) and (d) for the Mg$_4$, Ca$_4$, Sr$_4$, and Ba$_4$ tetramers at the regular tetrahedral geometry. Panels (a) and (c) present interaction energies $V_\text{int}$, while panels (b) and (d) their decompositions into the additive two-body part $V^{2b}$ and the non-additive three-body $V^{3b}$ and four-body $V^{4b}$ contributions.}
\label{fig:PESs}
\end{center}
\end{figure*}

The trimer-formation chemical reactions~\eqref{eq:R2} are highly exoenergetic because of relatively large binding energies of corresponding trimers stabilized by non-additive three-body interactions. The characteristics of alkaline-earth-metal triatomic molecules in the singlet $^1A_1$ electronic ground state at the isosceles (for heteronuclear $A_2B$ combinations) and equilateral (for homonuclear $A_3$ ones) triangular equilibrium geometries are collected in Table~\ref{tab:A2B}, where the equilibrium angles between the two even legs of the triangle $\alpha_e$, equilibrium leg lengths $R_e$, well depths $D_e$, and non-additive three-body part of the binding energies $D_e^{3b}$ are presented.

A strong non-additive three-body contribution to the binding energy is observed for all alkaline-earth-metal combinations. This effect is most pronounced when two beryllium atoms are present. For Be$_2$Mg, Be$_2$Ca, Be$_2$Ba, Be$_2$Ra, and Ra$_2$Be, the non-additive three-body interactions are larger than the corresponding binding energies. It means that these molecules have repulsive two-body interactions already at their equilibrium geometries which occur at distances smaller than inner classical turning points of corresponding dimer potentials. The equilibrium distances for all trimers are shortened by around 0.5-1 bohr as compared with equilibrium distances in corresponding dimers, and they increase with increasing atomic numbers. All equilibrium angles are acute and larger (smaller) than 60 degrees if $A$ atoms are larger (smaller) than a $B$ atom in $A_2B$.

Figure~\ref{fig:PESs} presents the interaction energies in panel~(a) and their decomposition into additive two-body $V^{2b}$ and non-additive three-body $V^{3b}$ contributions in panel~(b) for homonuclear alkaline-earth-metal trimers (Mg$_3$, Ca$_3$, Sr$_3$, and Ba$_3$) at the equilateral triangular geometry as a function of the internuclear distance. The non-additive three-body contributions for the studied trimers are attractive in the large range of configurations, and their magnitude and stabilizing effect increase significantly with decreasing internuclear distance. At the equilibrium geometry, the non-additive three-body contribution is about 96\% of the interaction energy for Mg$_3$ and is less important but still significant for other trimers. At larger distances, the additive terms start to dominate the interaction energy because they decay slower than non-additive parts ($\sim{1}/{R^6}$ vs.~$\sim{1}/{R^9}$). Finally, the importance of non-additive stabilization slightly decreases with increasing atomic numbers because of increasing additive contributions. 

Figure~\ref{fig:PESs} also presents the interaction energies in panel~(c) and their decomposition into additive two-body $V^{2b}$ and non-additive three-body $V^{3b}$ and four-body $V^{4b}$ contributions in panel~(d) for homonuclear alkaline-earth-metal tetramers (Mg$_4$, Ca$_4$, Sr$_4$, and Ba$_4$) at the regular tetrahedral geometry as a function of the internuclear distance. The studied tetramers are relatively strongly bound. Interestingly, the non-addictive four-body contributions are repulsive in the studied range of configurations, and tetramers are stabilized solely by the non-addictive three-body interactions, similarly as already observed for similar alkaline-earth-metal~\cite{MatoJCP2022} and spin-polarized alkali-metal~\cite{TomzaPRA13} clusters.

\section*{Experimental consequences}

Our results show that bimolecular atom-exchange and atom-transfer chemical reactions are energetically allowed between all ground-state alkaline-earth-metal diatomic molecules. It means that these reactions proceed with releasing energy, where a part of the chemical binding energy of reactants is converted into the kinetic energy of products allowing for trap losses. Additionally, there are no electronic barriers on intermolecular potential energy surfaces to prevent such reactions. Therefore, ultracold gases of alkaline-earth-metal diatomic molecules are chemically unstable with respect to bimolecular collisions, and loss rate constants close to the universal limit can be expected~\cite{IdziaszekPRL10}. These chemical reactions, controlled with external electromagnetic fields, can be an interesting subject of study~\cite{LeungNJP21}. However, for practical application of such molecules, chemical losses should be suppressed using either segregation in an optical lattice~\cite{StellmerPRL12,MosesScience2015} or shielding schemes with microwave~\cite{KarmanPRL18,LassablierePRL18} or optical~\cite{XiePRL20} fields that were recently experimentally demonstrated~\cite{MatsudaScience2020,AndereggScience21,SchindewolfNature22}.

Similar conclusions should also be true for chemical reactions between isoelectronic molecules based on alkaline-earth-like ytterbium~\cite{TomzaPCCP11}, zinc~\cite{ZarembaPRA21}, cadmium~\cite{ZarembaPRA21}, or mercury atoms. Additionally, molecules containing highly magnetic lanthanide atoms such as Dy~\cite{FryePRX20}, Er~\cite{TiesingaNJP21}, or Eu~\cite{TomzaPRA14} may present a similar pattern of chemical reactivity and strong non-additive three-body interactions because of their $4f^n6s^2$ valence electron configuration, where the $4f^n$ inner open shell is chemically screened by the $6s^2$ outer closed shell. Detailed study of such molecules is, however, out of the scope of this work.

\section*{Conclusions}
In this Letter, we have studied the energetics of both atom-exchange and trimer-formation chemical reactions between ultracold alkaline-earth-metal diatomic molecules in their absolute ground state. We have shown that both types of reactions are energetically allowed and barrierless for all alkaline-earth-metal diatomic combinations. Consequently, ultracold gases of such molecules can be a subject of fast reactive trap losses~\cite{IdziaszekPRL10}, in agreement with the recent experimental observations for ground-state Sr$_2$ molecules~\cite{LeungNJP21}. To predict the energies of trimers, we have performed high-level electronic structure calculations on the potential energy surfaces of all alkaline-earth-metal triatomic molecules and have reported their equilibrium geometries and dissociation energies. We have found that the alkaline-earth-metal trimers are stabilized by the strong non-additive three-body interactions. Finally, we have studied the non-additive three-body and four-body interactions in selected alkaline-earth-metal tetramers, which also are stabilized by the strong non-additive three-body contributions.
 
The predicted chemical reactivity of the ground-state alkaline-earth-metal diatomic molecules is both a challenge and an opportunity. The quantum-controlled and state-resolved experimental studies~\cite{HuScience19} of the interplay of the atom-exchange and atom-transfer reactions, as well as photoinduced losses~\cite{ChristianenPRL19}, can give new insights into the physical basis of cold chemistry. The use of different bosonic (spinless) and fermionic (spinful) isotopes can be another control knob~\cite{TomzaPRL15} to probe and understand long-lived intermediate four-atom complexes in ultracold bimolecular collisions~\cite{JachymskiPRA22,Man2022}. Additionally, chemical reactions can be avoided in ultracold mixtures of alkaline-earth-metal atoms and molecules, which can serve as a testbed for understanding chemical reactivity in more complex systems~\cite{JachymskiPRA22}.

\begin{acknowledgements}
Financial support from the National Science Centre Poland (grant no.~2020/38/E/ST2/00564) and the European Union (ERC, 101042989 -- QuantMol) is gratefully acknowledged. Views and opinions expressed are however those of the authors only and do not necessarily reflect those of the European Union or the European Research Council. Neither the European Union nor the granting authority can be held responsible for them. This research was supported in part by the National Science Foundation (Grant No.~NSF PHY-1748958). The computational part of this research has been partially supported by the PL-Grid Infrastructure (grant no.~PLG/2021/015237).
\end{acknowledgements}

\bibliography{EAM_dimers}

\end{document}